# Routing Proposals for Multipath Interdomain Routing

Sardar M. Bilal[1], Muhammad Naveed Dilber[2], Atta ur Rehman Khan[3]
[1]Department of Telematic Engineering,
University Carlos III de Madrid, Madrid, Spain.
[2] Department of Computing,
SZABIST, Karachi, Pakistan.
[3]Faculty of Computer Science & Information Technology,
University of Malaya, Kuala Lumpur, Malaysia.

sbilal@it.uc3m.es, naveed.dilber@szabist.edu.pk, attaurrehman@siswa.um.edu.my

*Abstract*— **Internet is composed of numbers of independent autonomous systems. BGP is used to disseminate reachability information and establishing path between autonomous systems. Each autonomous system is allowed to select a single route to a destination and then export the selected route to its neighbors. The selection of single best route imposes restrictions on the use of alternative paths during interdomain link failure and thus, incurred packet loss. Packet loss still occurs even when multiple paths exist between source and destination but these paths have not been utilized. To minimize the packet loss, when multiple paths exist, multipath routing techniques are introduced. Multipath routing techniques ensure the use of alternative paths on link failure. It computes set of paths which can be used when primary path is not available and it also provides a way to transit domains to have control over the traffic flow. This can be achieved by little modification to current BGP. This paper highlights different multipath routing techniques and also discusses the overhead incurred by each of them.**

## I. INTRODUCTION

Internet is composed of numbers of independent autonomous systems (AS). An AS exchange information in two different manners; Interior Gateway Protocols (IGP) and Exterior Gateway Protocols (EGP). IGP, the intradomain routing protocol, runs within single AS to disseminate local information whereas EGP, interdomain protocols, runs across different ASes to distribute reachability information. Border Gateway Protocol (BGP) is the only protocols that employed on Internet for disseminating reachability information and selects routes according to local policy. BGP selects the single best route towards each destination prefix. This provides limited control over the traffic flow, for instance, AS A wants to avoid AS E but cannot do so because both AS B and AS D have selected their best route through AS E.

BGP is a policy-based protocol which selects routes based on commercial incentives rather than selecting the shortest AS-path route. The router decides which route to disseminate to its neighbors. A business relationship determines selection and dissemination of route. The business relationships are customer-provider, peering, and sibling [1, 4]. In customer-provider relation, customer pays for the transit services to its provider. The provider in return, advertises the routes of its customer to its neighbors. Customer only announces the route of its provider to its own customers. In peering, two ASes agree to exchange traffic between each other's customers free of charge. Peering routes are only advertised to customers. Sibling ASes are the part of same institution, for example large ISP. Sibling ASes provide transit services to one another. If an AS has multiple routes towards the same destination, an AS prefers to route traffic on customer-learned routes, then route learned from sibling, then peers, and in the last it prefers to use route learned from its providers. This routing policy is called prefer-customer. When customers are not allowed to transit traffic from one provider to another and peers are not allowed to transit traffic from one peer to another, such type of routing policy is known as valley-free policy.

In current interdomain routing system, each router selects a single best route towards each destination prefix among different available routes and then advertises a single route to its neighbors. Thus, alternative routes are not announced. For example, in fig 1, AS B has learned the route B-C-F through AS C but it will not announce it to AS A. The current system is not flexible enough to take advantage of multipath if a failure occurs. Similarly, current system will provide very little control over the traffic flow. AS A wants to send its traffic to AS F via link B-C. The current system will not allow this circumvent. In current system, the selection of routes depends upon business relationships between ASes. In fig 1, AS B prefers route B-E-F due to financial reason but it may be willing to use route B-C-F to send AS A's traffic. Current system provides very little control for one domain to influence other domain's choice. For example, AS F wants to balance its incoming traffic over the links C-F and E-F and for that it uses prepending technique to balance the traffic which can be nullified by other ASes local policy [5].

In this paper, we explain different multipath techniques for interdomain routing and explain flexible way that gives control to transit ASes over traffic flow through their infrastructure.

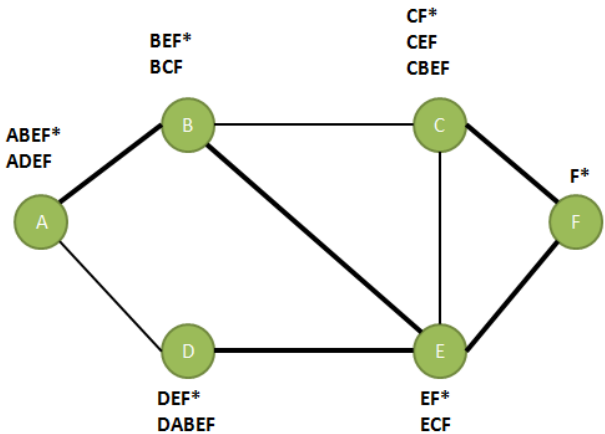

Figure 1: Single-path routing

In order to circumvent the AS, due to security reason or performance requirement, a multipath technique called Multi-path Interdomain Routing (MIRO) provides diversity in path selection, avoid state explosion, and give intermediate ASes control over the traffic that passes through their network. The detailed working mechanism of MIRO is presented in section III. Multipath techniques are designed to overcome disconnectivity problem and ensures continues connectivity between source and destination if alternative path exist. Whenever a link goes down, the packet should be forwarded around that link but BGP takes time in order to find other suitable path that avoids the link. The time BGP takes to find alternative path causes a transient disconnectivity and packets are dropped during that period. Transient disconnectivity will be explained in section IV. The routing protocols which handle transient disconnectivity are yet another Multipath Routing scheme (YAMR) and Resilient BGP (R-BGP). Both techniques provide alternative path when link goes down or path is unavailable, YAMR computes alternative path against each link of primary path or default path whereas R-BGP pre-computes alternative paths called failover path. Both will be discussed in more detail in section V and VI respectively and finally we conclude in section VII.

## II. RELATED WORK

It has been shown that BGP lost packets during convergence. Whenever a link goes down, the packets are forwarded around that link but BGP takes time in order to find other suitable path for packet forwarding. The time BGP takes to find alternative path, causes a transient disconnectivity and packets are dropped during that period. On average, 30% of packet loss within two minutes due to route change [6]. Hundreds of loss burst can be generated by a single routing event, out of which some last up to twenty seconds [7]. Due to BGP dynamics, the router suffers transient disconnectivity [8, 9]. The topology of Internet is considered highly redundant [4, 5] and BGP still suffers from transient disconnectivity even when there are alternative paths available from source to destination. Multipath techniques are designed to overcome disconnectivity problem and ensures continues connectivity between source and destination if alternative path exist. Previously, research was focus on minimizing the convergence time [10, 11] which ultimately increase the BGP complexity. Considering the size of Internet, fast convergence is very difficult to achieve especially for real time applications. Real time applications cannot tolerate more than couple of seconds of disconnectivity. So, fast convergence is difficult to achieve for such kind of applications.

In the past, numerous researchers work on providing flexible path selection mechanism and proposed source routing [15, 16, 17]. In source routing, source builds the list of intermediate routers through which packet has to travel to reach destination. Source routing provides flexibility in path but suffer from several issues. The most vital issue is that the intermediate routers have very little amount of control over traffic flow and thus, cannot engineer their network traffic. Secondly, as source routing computes complete end-to-end route to the destination, the source router should have complete knowledge of network topology. Computing end-to-end path incurs large overhead. On link failure, the source must have the knowledge of new topology which affects the stability and efficiency [14].

Recently, researchers divert their attention to multipath routing techniques, in which source selects alternative path if preferable path is not available due to link failure. Selecting a path among multiple available paths provide flexibility in term of user needs and provide control over the traffic flow. Designing of intradomain multipath is very difficult, but interdomain is much more complicated due to presence of policy constraints and scalability issues. There are some multipath routing protocols [5, 20] that address these two issues and shows significant performance improvement.

## III. MULTIPATH INTERDOMAIN ROUTING

MIRO is an Interdomain routing protocol which offers substantial flexibility, avoid state explosion, give intermediate ASes control over the traffic that passes through their network, and backward compatible. It is divided into five parts [5]; AS-level path selection, pull-based route retrieval for scalability with no additional overhead, bilateral negotiation between ASes, selective export of extra routes in order to obtain control of traffic traversing the network, and tunneling for forwarding data packets towards destination.

### A. AS-Level Path Selection

In conventional BGP, the route advertisement is propagated to a neighboring AS after AS adds its own AS number to the AS-path attribute. In source routing, path is selected at link-level which exposes the links of the intermediate ASes. Infect, this makes it less scalable than AS level path selection when link failure occurs in topology.

MIRO is based on AS level path selection which allows intermediate ASes to hide their internal structure, thus, increase scalability by giving control to ASes.

*B. Pull Based Route Retrieval*

Like conventional BGP, MIRO propagates default route to every neighbor but it cannot propagates alternatives routes to every neighbor because by doing so, it can limit the scalability. MIRO only propagates the alternate routes on demand. For example, if AS A does not want its traffic traversing AS E moving towards the destination AS F, then it can request the AS B to propagates the alternative route if any, towards AS F without traversing AS E while other ASes use default route. Pull based route retrieval not only allows ASes to propagate alternate routes on demand but it also allows other ASes which have not deployed MIRO work the same way as before. For instance, AS C and AS F do not need to deploy MIRO routing protocol while AS A requests alternative routes from AS B.  Each AS independently decides whether to use MIRO protocol to provide added services to their customers.

*C. Bilateral Negotiation Between ASes*

Mostly business relationships are bilateral. In bilateral negotiation, an AS initiates the negotiation as a requesting AS and other AS responses as a responding AS. The AS which generates the packet or initiates the negotiation is called upstream AS and which receives the packet or responding AS is called downstream AS. AS A which is upstream and also the requesting AS initiate the negotiation by sending the request to its neighbors, i.e., AS B and AS D of alternative paths towards AS F that avoids AS E. If alternative path is available then downstream AS will send to the upstream AS (AS A), otherwise AS B may ask AS C about the path that avoids AS E. This provides greater flexibility when neighbors ASes have not deployed the MIRO routing protocol. In this scenario, AS A requests alternative paths from AS C as A's neighbors have not deployed the MIRO routing protocol. It can also be helpful in traffic engineering, for example, the link C-F is overloaded by the traffic send by AS A, AS B, AS C, and AS E, then AS F request the more than one AS to redirect their traffic that uses the  link C-F.

*D. Selective Export of Extra Route*

The responding AS may have more than one path that fulfills the requirement of requesting AS but it cannot propagate all the paths that fulfill the requirement. This may cause considerable overhead. Responding AS propagates paths based on its routing policies and also tags routes with pricing information in order to let the requesting AS to decide which path to follow. For example the customer of AS C wants to avoid link C-F, AS C announces routes C-E-F and C-B-E-F with pricing information because traversing AS B might cause financial cost. The responding AS would prefer customer learn paths rather than paths learned from its peers or providers.

*E. Tunneling for Forwarding Data Packets*

After successful negotiation between two ASes, both create tunnel for data forwarding. The responding AS also provides the unique identifier to the requesting AS. After successful negotiation between AS A and AS B, AS B allocates the ID 7 and forwards it to AS A. AS A forwards the packets into tunnel and AS B remove the packets from tunnel and send towards the AS C via (B-C) link. AS C will then forwards the packets towards AS F by using destination IP address. AS A directs real time traffic via link (C-F) while send other low priority traffic through link (E-F). If the path B-C-F is no longer available then AS B tears down the tunnel and if path A-B changes then AS A tears down the tunnel.

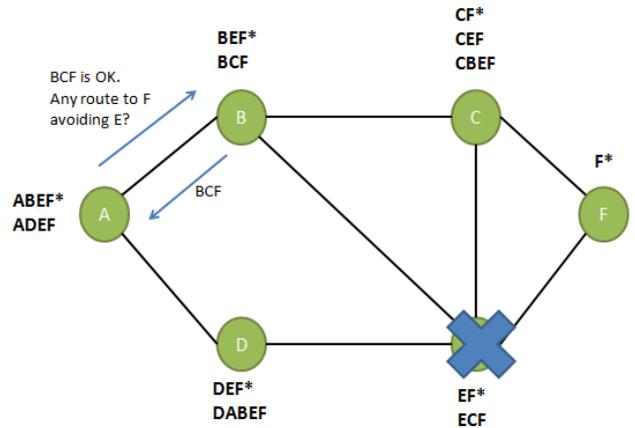

Figure 2: Route negotiation

IV. TRANSIENT DISCONNECTIVITY PROBLEM

Whenever a link goes down, BGP takes time to find other suitable path that avoids the disconnected link.  That is one of the reasons why we need a multipath situation in case of disconnectivity. The time BGP takes to find alternative path causes a transient disconnectivity and packets are dropped during that period. Packet loss causes delays as well as sender has to retransmit the data. In figure 3(a), MIT is a customer of both Sprint and Hari. Hari is a customer of AT&T and Peter.

AT&T and Peter will not announce the path towards MIT, since Hari advertises the path to reach the MIT. As AT&T and Peter do not announce any path to Hari, Hari has no alternative path to which it can switch the traffic when the link between Hari and MIT fails. As Hari has no alternative path so it drops all the packets destined for MIT. It not only drops its own traffic but it also drops the traffic of its upstream (AT&T and Peter) [21] as shown in figure 3(b).

Hari withdraws the path to MIT illustrated by figure 3(c) AT&T and Peter moves to alternative path and soon Hari resumes its traffic when AT&T announces alternative path to MIT through Sprint as show in figure 3(d). This temporary disconnectivity is known as transient disconnectivity. Hari drops all the packets destined to MIT during transient disconnectivity, even there is an alternative path available via Sprint. Whenever a routes changes, it causes transient disconnectivity which might takes few minutes [28, 7]. This delay includes the searching of upstream AS which has alternative path and then search the path which go around the link. AT&T prefers customer path i.e., a path from Peter to MIT via Hari. But this path includes the link which is not available (Hari to MIT). Thus, AT&T discards this route and announces a path towards MIT via Sprint. Finally, link failure generates a number of update messages for all destination prefixes that uses the link [22].

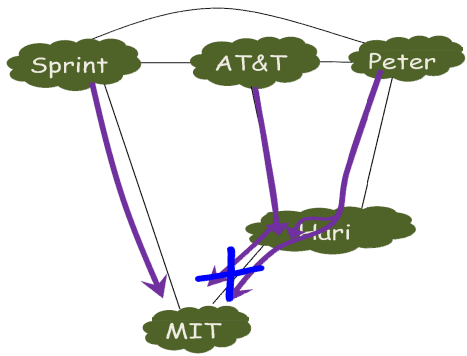

Figure 3(a): Link failure Hari->MIT

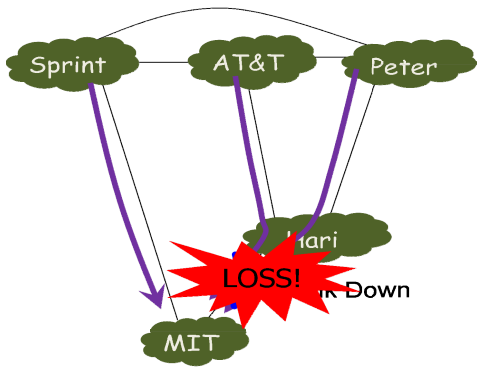

Figure 3(b): Hari, AT&T, and Peter packets are dropped

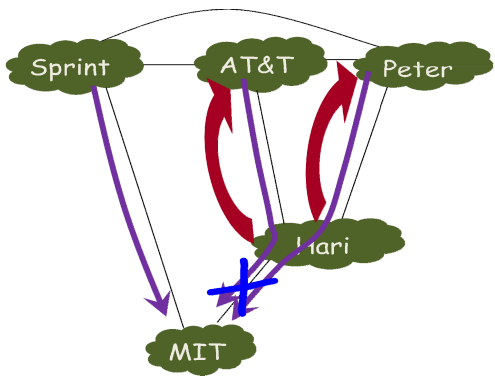

Figure 3(c): Hari withdraws path

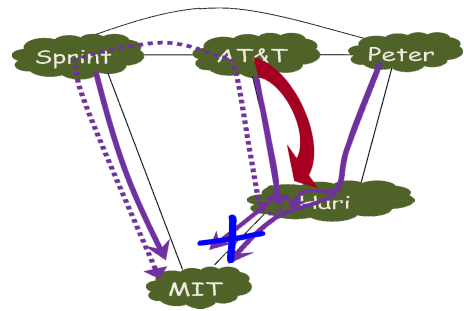

Figure 3(d): AT&T announces path to Hari via Sprint

## V. RESILIENT-BGP

Due to huge size of Internet, announcing alternative failover paths may lead to explosion of routing messages, even announcing a single failure path to each neighbor would require too much exchange of routing messages, making the system unscalable. In R-BGP, AS announces one failover route to one strategic neighbor. R-BGP ensures continuous connectivity even in the case of link failure by pre-computing failover paths. The real challenge lies in the transformation from failover state to converged state. This could cause routing loops or making the domain inconsistent by making believe that there is no available route to the destination, even the route does exist. To prevent inconsistent state, R-BGP disseminates extra information in update message.

The concept of failover path has been used previously in MPLS fast re-routing [23], and managing second recovery on link failure within AS [24, 25]. This work consider only intradomain routing, interdomain routing has its own challenges like AS policies, routing metrics, etc. In interdomain routing, failover path is used as a way of providing multiple BGP paths [5], while R-BGP use failover paths to provide continue connectivity. In conventional BGP, AS advertises only its preferable path to the neighbors.

It advertises alternative path after a link goes down. The main idea of R-BGP is to advertise the alternative path (failover) before a link goes down. As shown in fig. 3(c), when a link between Hari and MIT goes down, Hari finds no alternative path to forward the packets towards MIT [21]. In this case, Hari drops the packets of AT&T and Peter and also drops its own packets destined for MIT until AT&T announces alternative path via Sprint as shown in fig. 3(d). In case of failover path, AT&T informs the Hari of alternative path towards MIT via Sprint before the link goes down. Wherever link between Hari and MIT goes down, Hari uses failover path to direct the traffic to AT&T. AT&T forwards the traffic to MIT via Sprint. Hari only uses failover path if there is no alternative path available from Hari to MIT. By using failover path, Hari will not drop its own packets in addition to Peter and AT&T packets.

There are two challenges associated with propagation of failover paths which are stated as under and both are of utmost importance [21].

- First, minimize the dissemination of numbers of failover path in order to control the overhead.
- Second, transition from stable path to failover path without creating loops.

*A. Selecting and Advertising Failover Path*

There are few questions which needs to be answered, for instance, which failover path should be advertised, and to whom failover path should be announced. BGP announces one path per destination to its neighbors. Announcements of multiple paths to each neighbor will results in explosion of routing messages which incurs large overhead and as well as affect the scalability. Just like BGP, announces at most one failover path to each neighbor. If neighboring AS don't have the failover path, then it will drop all the packets of its upstream ASes. So, if an AS immediate above the down link knows the failover path, the AS will divert the traffic on failover path when the link fails. Each AS will announce one failover path and it will announce only to the AS through whom it is routing (i.e., Hari) as shown in figure 4(a). Peter don't receive any failover path from AT&T, So, when link between Hari and MIT goes down, Peter will forwards the packet towards Hari as before until Peter receive advertisement from AT&T. The key idea is to forward the packets towards destination irrespective of what happened at the routing layer.

Secondly, which failover path it needs to announce? The answer is to announce the path which is most disjoint path from the primary path, so that overlapping link or the path which include the failed link cannot be used again (failed link can be used if we consider second best route). For example, in fig. 4(c), the David most disjoint path is the one which it learns from AT&T and then forward this path to Hari. If the link between Hari and MIT goes down, Hari will divert its traffic towards AT&T via David. Disjointness can be measured on the basis of common suffix. The disjointness path is the one at upstream AS which does not include down link because down link has a longer shared suffix. The longer the common suffix, the more is the possibility that failover path may contain the failed link.

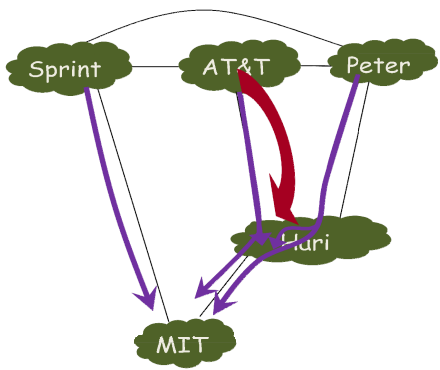

Figure 4(a): AT&T announces failover path to Hari

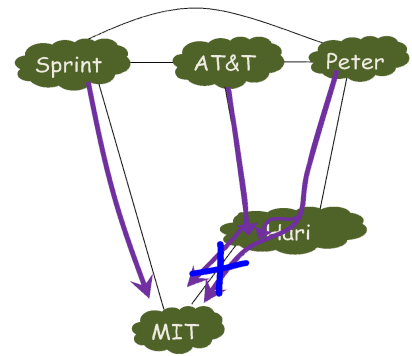

Figure 4(b): Link failure Hari-> MIT

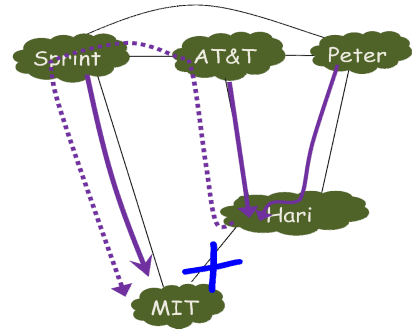

Figure 4(c): Hari send traffic on failover path

*B. Avoiding Routing Loops*

To ensure continuous connectivity between ASes, either in the case of link failure or because of routing loops, transition from link failure to failover path must avoid the routing loops. For example, if the link between Hari and MIT goes down, Hari divert its traffic to AT&T and withdraws its route to MIT and send route withdraw information to AT&T and Peter.

According to the policies of BGP, Hari does not allow to advertise the path from one provider (AT&T) to other provider (Peter). The route withdraw announcement does not determine the reason of withdraw, which allow Peter to sends its traffic to MIT via AT&T and Hari while AT&T sends traffic to MIT via Peter and Hari. This cause routing loops which BGP fix later but until then all the packets will be dropped.

Routing loops could be avoided by informing both AT&T and Peter that link between Hari and MIT goes down, So don't forward the packets that uses the same link. This can be achieved by introducing Root Cause Information (RCI) [21] which can be added in withdrawal advertisement. RCI make sure that new path is loop free and it stops other ASes of traversing the same down link.

R-BGP shows that by using most disjoint paths can avoid transient disconnectivity during BGP convergence. R-BGP uses failover paths to forward the traffic on alternative path on link failure. The undue overhead in failover path dissemination is controlled by advertising at most one failover path to next AS which belongs to its primary path.

## VI. YET ANOTHER MULTIPATH ROUTING

Yet Another Multipath Routing protocol systematically provides high path diversity. There are two components to YAMR [27]; diverse paths calculation and churn reduction technique. YAMR Path Construction (YPC) computes alternative path for each link of default path, so that if the link goes down in default path, the traffic can be rerouted through alternative path. Thus, no single failure can break all the paths. The computation of diverse paths involves higher number of messaging overhead. YAMR introduces new distributed mechanism to hide a failure in a path by not propagating to all nodes. This technique makes the YAMR more resilience.

### A. YAMR Path Construction

The main idea of YPC is to construct a default path and then construct an alternative path for each link of default path avoiding that link. Default path is computed the same way as BGP computes the path. Default as well as alternative paths are computed from the neighbors advertised paths. Paths learnt from the neighbors are store in RIB_IN whereas RIB_LOCAL contains the set of paths used for forwarding.

Labels are used to make distinguish between multiple paths towards the same destination. Default paths are identified by d label whereas alternative paths are labeled by the link they avoid [27].

Figure 5 helps to understand the path computation and path selection mechanism of YAMR. Initially, AS D announces its default path to AS C and AS E. Then AS C and AS E announce the path to each other. On exchanging the default path to each other, C and E compute alternative path. If AS C uses [C, D] as default path so, on failure of the link (C-D), C will use alternative path [C, E, D]. Then C sends the updates of its LOCAL_RIB to B and F which send their LOCAL_RIB to A after calculating their alternative paths. AS A construct its default path and alternative paths, for instance, AS A would like to use [A, B, C, D] as its default path, from LOCAL_RIB updates from AS B and AS F.

On receiving update of LOCAL_RIB from F and B, A can compute alternative path for each link of its default path [A, B, C, D]. AS A computes alternative path [A, F, C, D] that avoids the link (A-B). After that, A computes the alternative path that avoids the link (B-C); again the computed path is [A, F, C, D]. Finally, A computes the alternative path [A, B, C, E, D] that avoid the link (C-D).

If AS A receives a packet for destination D, then AS A will forward the packet by using its L-labeled path (one that exists in its LOCAL_RIB) if the packet is labeled with L, otherwise it uses its default path to forward the packet towards D. In case, AS A doesn't have a default path towards AS D, the packet is dropped. For each destination, YAMR computes a default path and then computes alternative path against each link of default path. It means it will take more control messages than BGP which will increase overhead. YAMR uses hiding churn reduction technique to reduce the overhead.

### B. YAMR Hiding Technique

The distributed hiding technique can be applied to both YAMR and as well as BGP. The core idea of this technique is to propagate the link failure to subset of neighbors instead of all propagating it to all neighbors. Hiding AS will not propagate the link failure to the neighbors if it can reroute the packet avoiding the failed link. For example in figure 6(a), if link (N-D) fails, AS N can still reroute the packet through [N, O, D] avoiding the failed link (N-D) without telling AS I, AS G and AS L of the link failure. In this case, AS N is called hiding AS, the path [N, O, D] is reflection path and path [N, D] is called lame path [27].

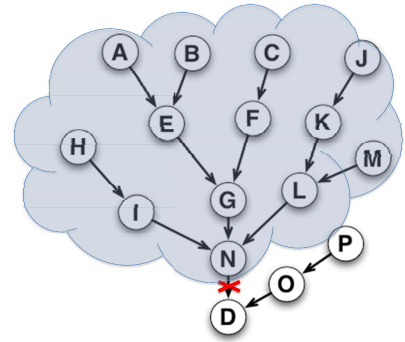

Figure 6(a): Link failure

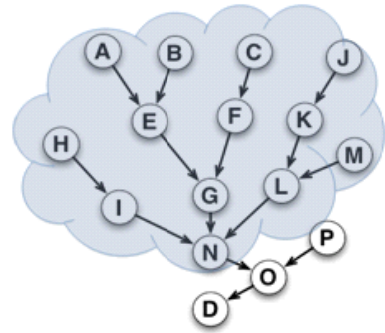

Figure 6(b): Failure hiding at AS N

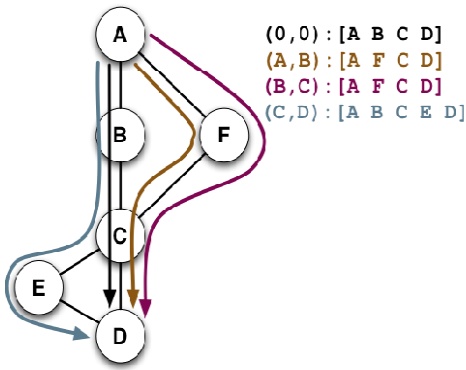

Figure 5: YAMR Path Construction

It is not always the case, that AS N hides the path to all its neighbors. It can't hide the path to the neighbor to whom it exports its suitable path. So, AS N withdraws the path from them. These neighbors then hide failure from their neighbors and it continues until failure is completely hidden. The link failure occur between AS N and AS D, allows AS N to hide this failure by pretending that the lame path is available, so when path is withdrawn, N will not delete it from its RIB_IN unlike the BGP. AS N called this a lame path and tries to route the traffic onto a deflection path if it has one, for instance, in this case N reroutes along its alternative path [N, O, D] but does not change its advertised path [N, D]. If there is no alternative path, the hiding mechanism stops and the path is deleted from the RIB_IN. The steps involved in hiding failure mechanism are:

- Don't delete a withdrawn path, mark it lame.
- For each selected lame path, pick a deflection path.
- If unsuccessful, treat as a normal withdrawal, delete the lame path and redo selection.
- Advertise the lame path, but route on deflection path.

## VII. CONCLUSION

This paper explains the importance of multipath technique and explains several state of the art multipath techniques. Multipath techniques allow routers to divert traffic to alternative path on link failure. The basic objective of this work is to show whether multipath techniques can be implemented to current network architecture while keeping the same overhead as that of BGP and provide scalability, reliability, and avoid routing loops.

Multipath techniques like MIRO, R-BGP, and YAMR route the packets to alternative path when a link goes down. MIRO is capable of circumvent AS due to performance or security reason and provides mechanism for bilateral negotiation between ASes. In data plane, it uses tunnel to forward traffic. R-BGP provides resilience against link failure by offering failover path i.e., the one which is most disjoint path from the primary path, and avoids loops in the network by introducing RCI. There will be no packets drop by using failover path during transient disconnectivity. YAMR computes alternative paths on each link of the primary path in order to avoid packets drop during link failure. This sound that there would be possible a large overhead involved in calculating and disseminating of alternative paths against each link on primary path but YAMR uses hiding technique to hide the failure from the upstream if it has alternative path, thus reduce the overhead.